\documentstyle[prl,aps,preprint]{revtex}
\begin{document}
%\tightenlines
\draft
%\twocolumn
\widetext

\title{On the relation of the gravitino mass and the GUT parameters}
\author{ V.I. Tkach$^a$ \thanks{E-mail: vladimir@ifug1.ugto.mx},
J.J. Rosales $^{a,b}$\thanks{E-mail: juan@ifug3.ugto.mx}, 
and J. Torres $^a$\thanks{E-mail: jose@ifug4.ugto.mx}\\ 
$^a$Instituto de F\'{\i}sica de la Universidad de Guanajuato,\\
Apartado Postal E-143, C.P. 37150, Le\'on, Guanajuato, M\'exico\\
$^b$ Ingenier\'{\i}a en computaci\'on, Universidad del Baj\'{\i}o\\
Av. Universidad s/n Col. Lomas del Sol, Le\'on, Gto., M\'exico}

\maketitle
\date{\today}

\begin{abstract}
In this article we consider the local supersymmetry breaking and the 
broken SU(5) symmetry permisible by dilaton vacuum configuration in 
supergravity theories. We establish the parameter relation of spontaneuos 
breaking of supersymmetry and of the GUT theory. 

\end{abstract}
\pacs{PACS: 04.65.+e, 11.30.Qc, 12.60.Jv.}
\narrowtext 
\newpage

The structure of the effective N=1 supergravity theory cannot provide
the small vacuum energy, this explanation may be only under the
construction of the quantum gravity theory, possibly in the frame of 
superstring theories. The question about spontaneous supersymmetry 
breaking and the arising of small scale mass $10^2 \sim 10^3$ Gev 
probably can be determined from the supergravity theory or from their 
effective supergravity theory, these possibilities are intensively 
discussed in the literature. Quadratically divergent one-loop corrections 
to finite contributions of the effective potential $ m_{3/2}^2 M^2_{pl}$ 
could destabilize the hierarchy $m_{3/2}<<M_{pl}$. Moreover, the 
$m_{3/2}^2 M^2_{pl}$ contributions to the vacuum energy cannot be cancelled 
by symmetry breaking phenomena ocurring at much lower energy scales. 
Possibly this realization will be understood in the future theory with 
hidden symmetry leading to the vacuum energy $m_{3/2}^2 M_{pl}^2$ 
elimination \cite{uno}. Nowadays the most natural candidate for such 
theories is the heterotic superstring \cite{dos}.    

In all the models of spontaneous breaking of local supersymmetry an 
additional intermediate mass scale $M_{hidd}$ is introduced, so that 
$M_w <<M_{hidd} << M_{pl}$ in order to have mass value for the gravitino 
in the range $10^2 \sim 10^3$ Gev, acceptable from the point of view of 
phenomenology, as well as from the point of view of quadratical divergences 
in the action for the absent fields in the minimal supersymmetric 
extension of the standard model (MSSM). While $M_{hidd}$ is a new mass 
scale coinciding with $M_{GUT}$ for the geometrical hierarchy case 
$10^{10}$ Gev \cite{{tres},{cuatro}}, the effective string theory with 
observable sector $SU(3)\otimes SU(2)\otimes U(1)$ and the hidden mass 
scale sector $10^{13}$ Gev \cite{cinco,seis,siete,ocho} are not coincided 
with $M_{GUT}$.

The problem of the vanishing vacuum energy in the classical level is 
determined by the so-called no-scale supergravity models \cite{siete}, 
however there is not stable minimum of the vacuum without flat direction 
\cite{{cinco},{siete},{nueve}}. In the case of the effective $N = 1$ 
supergravity theories we have dificulties with the supersymmetry breaking 
in the moduli direction (hidden sector) in the minimum of the potential 
\cite{{nueve},{diez}}.

In our previous works \cite{{once},{doce}} it was shown, that for 
spatially homogeneous part of fields in the supergravity theories 
interacting with matter fields there is a vacuum configuration invariant 
under $n=2$ local conformal supersymmetry. This supersymmetry is a subgroup 
of the four-dimensional space-time supersymmetry. As the requirements of 
the local conformal supersymmetry are not so hard as the requirements of 
the space-time supersymmetry, then the new possibilities in research of 
spontaneous supersymmetry breaking arise.  

The purpose of this letter is to provide a mechanism, which naturally 
generates a small scale of the order $10^2 \sim 10^3$ Gev. It is done 
without additional intermediate mass scale parameter. In this case we 
will have a stable minimum of the potential with zero energy in the 
classical level corresponding to tree-level aproximations; not breaking 
SU(5), not breaking supersymmetry and two minima with supersymmetry 
breaking, and SU(5) breaking in the phases $SU(3)\otimes SU(2)\otimes 
U(1)$ and $SU(4)\otimes U(1)$ states. The gravitino mass in the state 
$SU(4)\otimes U(1)$ will have a value in two orders less than $m_{3/2}$ 
in the states with symmetry of the standard model, and the mass 
$m_{3/2}(SU(3)\otimes SU(2)\otimes U(1))$ is defined by a constant 
self-interaction $\alpha_{GUT}= \frac{1}{26}$ and the GUT mass 
$M_{GUT}$. The construction of the spontaneous breaking supersymmetry 
mechanism is related to the existence of vacuum states in supergravity 
and the effective theory of supergravity invariant under the local 
conformal supersymmetry (which is a subgroup of $d=4$, $N=1$ supergravity) 
\cite{doce}.

The effective scalar field potential in the local conformal supersymmetry 
corresponding to the potential is given by  $V_{eff} = V_F + V_D$ 
\cite{doce}. 

\begin{equation}
V_F = \frac{e^{\alpha G}}{\kappa^4} \lbrack 
\alpha^2 G_{\bar A} G^{\bar A D} G_D - 3 \rbrack,
\label{potencial}
\end{equation}
where $\kappa = \frac{1}{M_{pl}}$ and 
$M_{pl} = \frac{1}{\sqrt {8\pi G_N}} = 2,4.10^{18}$ Gev. is the reduced 
mass Planck. The scalar field 
potential (\ref{potencial}) depends on the real gauge invariant K\"ahler 
function $G(z^A, \bar z^{\bar A}) = K(z^A, \bar z^{\bar A}) + 
\log |g(z^A)|^2$, where $K(z^A, \bar z^{\bar A})$ is the K\"ahler 
potential whose second derivatives determine the kinetic terms for the 
fields in the chiral supermultiplets and $g(z^A)$ is the superpotential. 
$\alpha$ is an arbitrary parameter, which is not fixed by conformal 
supersymmetry, and as it will be shown in this work it plays the role of 
the dilaton coupling constant \cite{trece}. Derivatives of the K\"ahler 
function are denoted by $\frac{\partial G}{\partial z^A}= G_A $,
$\frac{\partial G}{\partial \bar z^{\bar A}}= G_{\bar A} $, and the 
K\"ahler metric is $G_{A \bar B}= G_{\bar B A} = 
K_{A \bar B}=K_{\bar B A}$. The inverse K\"ahler metric $G^{A \bar B}$, 
such as $G^{A \bar B}G_{\bar B C}= \delta_A^C$, can be used to define 
$G^A \equiv G^{A \bar B}G_{\bar B}$ and $G^{\bar B} \equiv 
G_A G^{A \bar B}$. In our notation repeated indices are summed, unless 
otherwise stated. Note, that in contrast with global supersymmetry the 
$F$-term part of the effective scalar potentential in (\ref{potencial}) 
is not positive semi-definite in general. Therefore, it allows to have 
spontaneously supersymmetry breaking with vanishing classical vacuum 
energy, unlike in global supersymmetry. 

In order to discuss the implication of spontaneous 
supersymmetry breaking we need to display the potential (\ref{potencial}) 
in terms of the auxiliary fields $F_A$ of the matter supermultiplets

\begin{equation}
V_F = \frac{1}{\kappa^2}F_A \bar F^{\bar A} - \frac{3}{\kappa^4} 
e^{\alpha G(z^A, \bar z^{\bar A})},
\label{F-term}
\end{equation}
where $F_A$ has the following form

\begin{equation}
F_A = \frac{\alpha}{\kappa}e^{\frac{\alpha}{2} G(z^A, \bar z^{\bar A})}
G_A(z^A, \bar z^{\bar A}).
\label{auxiliar}
\end{equation} 
The local supersymmetry is spontaneously broken if the auxiliary fields
(\ref{auxiliar}) of the matter supermultiplets get non-vanishing vacuum
expectation values. According to our assumption at the minimum the 
potential (\ref{F-term}) is $V_F(z^A, \bar z^{\bar A})=0$, but 
$<F_A> = F_A(z_0, \bar z_0)\not=0$ and, thus, the supersymmetry is broken. 
The meassure of this breakdown is the gravitino mass $m_{3/2}$, which in 
our case is given by \cite{doce}

\begin{equation}
m_{3/2} = \frac{1}{\kappa}e^{\frac{\alpha}{2}G(z_0^A, \bar z_0^{\bar A})}= 
e^{\frac{\alpha G(z_0, \bar z_0)}{2}} M_{pl},
\label{masa}    
\end{equation}
which depends on the vacuum expectation values $z_0^A = <z^A>$ of the 
scalar fields of the theory determined by the condition of minimum 
vacuum energy. For convenience in the following we shall also classify 
the fields as $z^A \equiv (S, z^a)$, where $S$ stands for the dilaton 
field, while $z^a$ for the spatially homogeneous chiral fields. So, the 
conditions for the accurately spontaneous supersymmetry breaking with 
vanishing vacuum energy at the classical level is very simple if we take 
$\alpha = \sqrt 3$

\begin{equation}
\frac{\partial V_F}{\partial z^a}(S, \bar S_0, z_0^a, \bar z_0^{\bar a}) 
=0, \qquad\qquad V_F(S_0, \bar S_0, z_0^a, \bar z_0^{\bar a})=0, 
\qquad\qquad F_S(S_0, z_0^a) \ne 0,
\label{condisiones}
\end{equation} 
where $S_0$, $z_0^a$ are the absolute minima. The first condition implies
the existence of a minimum, the second condition implies the vanishing
cosmological constant, and the non-vanishing F-term implies the 
spontaneously supersymmetry breaking. We take the K\"ahler function as

\begin{equation}
G(S, z^a)= -\log(S + \bar S) + \frac{\kappa^2}{2} \bar z^{\bar a} z^b + 
\log\lbrace |\frac{\kappa^3}{2} g(z^a)|^2 \rbrace.
\label{funcion} 
\end{equation}
After substitution of (\ref{funcion}) into (\ref{potencial}) the 
effective potential becomes
\begin{equation}
V_F(S, z^a) = \frac{e^{\alpha G}}{\kappa^4}\lbrack \alpha^2 G_{\bar S} 
G^{\bar S S} G_S + \alpha^2 G_{\bar z^{\bar a}} G^{\bar z^{\bar a} z^a}
G_{z^a} - 3\rbrack.
\label{efectivo}
\end{equation}
Now, we will consider the SU(5) theory. The scale where the unified 
gauge symmetry is broken is described by a mass parameter $M_{GUT}$. 
Hence, the minimal choice of a superpotential is written as

\begin{equation}
\tilde g(z^a) = \tilde g(\Sigma) = \frac{1}{3} Tr\Sigma^3 + 
\frac{M_{GUT}}{2} Tr\Sigma^2,
\label{trace}
\end{equation}
where $g(z^a) = \lambda \tilde g(z^a)$ and $\Sigma$ is the adjoint 
representation 24 of SU(5) and $\lambda$ is a self-interaction coupling 
constant. In our analysis of the broken SU(5) GUT we will 
consider only the part of the supersymmetric potential \cite{catorce} 
depending on 
$z^a = \Sigma_y^x \equiv \Sigma$, which is the adjoint representation 
24 of SU(5), as the minimum of the scalar fields potential is achieved 
when the vacuum expectation values of the other left-handed multiplets 
are vanished. We consider the case when $g(S, z^a) = g(z^a)$. The 
condition $\partial_{\Sigma}g(z) = 0$ shows, that there is a minimum of 
the potential $V(\Sigma)$ for the global supersymmetry inclusively in 
the presence of the D-term in the effective potential, if 
$(\Sigma_y^x)_{diag}$ possesses one of the following vacuum expectation 
values $<\Sigma>$:

\begin{equation}				   
(i)  0, \qquad\qquad\qquad (ii) \frac{1}{3}M_{GUT}(1,1,1,1,-4),
\qquad\qquad\qquad
(iii) M_{GUT}(2,2,2,-3,-3)
\label{values}
\end{equation}
and the vacuum energy values of all other components of $\Sigma_y^x$ are
zero \cite{{cuatro},{catorce}}. Thus, solution (i) does not break SU(5),
while (ii) breaks SU(5) gauge group into $SU(4)\times U(1)$, and 
solution (iii) breaks the gauge group into $SU(3)\times SU(2)\times U(1)$. 
The supersymmetric self-interaction fields $\Sigma$ are constructed in 
such form, that in the mass scale of the gran unification  $M_{GUT}$ the 
broken symmetry of SU(5) takes place.

The contributions of D-terms in the effective potential preserve their 
forms and for the local conformal supersymmetry have the standard form

\begin{equation}
V_D = \frac{1}{2\alpha}(Re f^{-1})^{ij}(G_a(T_i)^a_{\bar b}
\bar z^{\bar b})(G_{\bar c}(T_j)^{\bar c}_d z^d),
\label{D-term}
\end{equation}
the functions $f_{ij}$ in this case have the form 
$f_{ij} = \delta_{ij} S$, and in particular the gauge coupling constant 
of SU(5) is done by $g^{-2}_{GUT} = <S>$. In the analysis of the 
effective potential $V_{eff} = V_F + V_D$ the stationary points 
corresponding to the minimum of $V_D$ term can be ignored because of the 
condition $G_{z^a} =0$, and this permits the analysis only for the 
$V_F$ term.  

Deriving $G(S, z)$ with respect to dilaton field in (\ref{efectivo}) 
we get
\begin{eqnarray}
G_S &=& \frac{\partial G}{\partial \bar S} = - \frac{1}{S + \bar S},
\qquad G_{\bar S} = \frac{\partial G}{\partial S} = -\frac{1}{S + \bar S}, 
\qquad G_{S \bar S} = \frac{1}{(S + \bar S)^2}, \\   
G^{S \bar S}&=& (S + \bar S)^2, \qquad G_S G_{\bar S} G^{S \bar S}=1,  
\nonumber
\label{ese}
\end{eqnarray}
and after substituting them again into (\ref{efectivo}) the potential 
becomes

\begin{equation}
V_F = \frac{3}{\kappa^4} e^{\sqrt 3 G} G_{\bar z} G^{\bar z z}G_z. 
\label{pot} 
\end{equation}
We see, that if $G_{z^a} = 0$ for any $z_0= <\Sigma_0>$ then 
$V_F\equiv 0$,
while modification $\frac{\partial g(z)}{\partial z^{a}} =0$ in 
$G_{z^a} = \frac{\partial g(z)}{\partial z^{a}} + 
\frac{\kappa^2}{2} \bar z^a = 0$, which leads to small correction in 
vacuum value $<\Sigma>$ (\ref{values}).

So, we compute the condition of the stationary points in the dilaton 
direction, ${\it i.e}$ $\frac{\partial V_{eff}}{\partial S} = 0$, 
we obtain

\begin{equation}
\frac{\partial V_F}{\partial S} = \frac{3 \sqrt 3}{\kappa^4}
e^{\sqrt 3 G} \lbrace G_{\bar z} G^{\bar z z} G_z \rbrace G_S = 0,
\label{direction}
\end{equation}
and
\begin{equation}
\frac{\partial V_F}{\partial \bar S} = \frac{3 \sqrt 3}{\kappa^4}
e^{\sqrt 3 G}\lbrace G_{\bar z} G^{\bar z z} G_z \rbrace G_{\bar S} = 0.
\label{complejo}
\end{equation}
The conditions of the stationary points in the dilaton direction are
$\frac{\partial V_F}{\partial S} = \frac{\partial V_F}{\partial 
\bar S} =0$ and can be satisfied in two different ways: as 
$G_{z^a}\ne G_{\bar z^{\bar a}} \ne 0$, then $G_S = G_{\bar S} = 0$, 
and therefore $F_S =0$ and the supersymmetry is not broken in the 
dilaton direction, on the other hand, if $G_S \ne 0$, $G_{\bar S} \ne 0 $ 
and $G_{z^a} = G_{\bar z^{\bar a}} =0$ then, we have broken supersymmetry 
in the dilaton direction. Therefore, state with not broken SU(5) 
$\it i)$ (\ref{values}) although $G_S \ne 0$, as soon as 
$(\tilde g(z_0^a) = \tilde g(\Sigma_0)= 0)$ is equal to zero, and 
$F_S$ is defined by (\ref{auxiliar}), then $<F_S> = 0$ 
and the supersymmetry is not broken. The states with broken SU(5) into
$SU(4)\otimes U(1)$ and $SU(3)\otimes SU(2)\otimes U(1)$ and $<F_S> \ne 0$
have broken supersymmetry. Minimizing the Eq. (\ref{efectivo}) with respect 
to the chiral fields we have
\begin{equation}
\frac{\partial V_F}{\partial z}= \frac{3}{\kappa^4} e^{\sqrt 3 G} 
G_{\bar z}G^{\bar z z} \lbrack \sqrt 3 (G_z)^2 + G_{z z} \rbrack = 0,
\label{chiral}
\nonumber
\end{equation}

\begin{equation}
\frac{\partial V_F}{\partial \bar z}= \frac{3}{\kappa^4} e^{\sqrt 3 G} 
G^{\bar z z}G_z \lbrack \sqrt 3 (G_{\bar z})^2 + G_{\bar z \bar z} \rbrack
=0. \label{field}
\nonumber
\end{equation}
The minimization of the potential (\ref{efectivo})
requires that $G_{z^a} = 0$ and, therefore, in the classical level the 
energy is equal to zero. In the case when $G_{z^a} \ne 0$ and 
$G_{\bar z^{\bar a}} \ne 0$ there are conditions for 
(\ref{chiral},\ref{field}) with $V_F > 0$, therefore, we get      
\begin{equation}
\sqrt 3(G_{z^a})^2 + G_{z^a z^b} =0, \qquad\qquad 
\sqrt 3(G_{\bar z^{\bar a}})^2 + G_{\bar z^{\bar a} \bar z^{\bar b}} =0,
\label{ecuaciones}
\end{equation}
but in this case to find the stationary points corresponding to maximum 
we need to considerate the $V_D$ contribution (of the D-term) 
(\ref{D-term}) in the relations (\ref{ecuaciones}). Neglecting 
$\frac{M^2_{GUT}}{M^2_{pl}}$ corrections to the solutions of the equation 
$G_{z^a} =0$, included in
(\ref{values}) the three solutions for $\tilde g(z^a)$ in 
(\ref{trace}) are
\begin{equation}
\tilde g(z_0^a) = (0, \frac{10}{27}M^3_{GUT}, 5 M^3_{GUT}).
\label{super}
\end{equation}
The third solution corresponds to the SU(5) breaking into
$SU(3)\otimes SU(2)\otimes U(1)$ and the second solution corresponds to 
$SU(4)\otimes U(1)$ state. Then, in this case the gravitino mass in
$SU(3)\otimes SU(2)\otimes U(1)$ state is

\begin{equation}
m_{3/2} = \frac{1}{\kappa}e^{\frac{\sqrt 3}{2} K(S_0, z_0^a)}
|g(z_0^a)|^{\sqrt 3} = \frac{1}{(2 S_R)^{\frac{\sqrt 3}{2}}}
\left( \frac{5}{2}\lambda \right)^{\sqrt 3}
\left( \frac{M_{GUT}}{M_{pl}}\right)^{3\sqrt 3},
\label{graviton}
\end{equation} 
where $S + \bar S = 2 ReS = 2S_R$ and $ReS = S_R$. The gravitino
mass $m_{3/2}$ is not fixed at the classical minimum, but it is a 
function of the dilaton field $S$ parametrizing the flat dilaton 
directon with the conditions $G_S\ne 0$ and $G_{\bar S} \ne 0$ in our 
case. The value $S_0 = <S>$ in the minimum with $V_F = 0$ is not fixed 
by K\"ahler function. 

In order to have stable minimum in the dilaton direction we modify the 
superpotential $g(S, z^a) = g(S)\tilde g(z^a)$. The requirement of 
vanishing vacuum energy imposes a non-trivial constraints on the 
structure of dilaton sector of the theory

\begin{equation}
\Pi \equiv G_SG^S - 1 = 0,\qquad at \qquad S=S_0,
\label{constraint1}
\end{equation}
and the constraint 
\begin{equation}
\partial _S \Pi = 0,\qquad at \qquad S=S_0,
\label{constraint2}
\end{equation}
which is necessary to preserve the stationary points conditions 
$G_{z^a}=0$ in the observable directions. In this case we have the 
conditions following from (\ref{constraint1})

\begin{equation}
\partial_S g(S) \partial_{\bar S} \bar g(\bar S) = 
\frac{g(S) \partial_{\bar S} g(\bar S)}{S + \bar S} + 
\frac{\bar g(\bar S) \partial_S g(S)}{S + \bar S},
\label{condisiones}
\end{equation} 
these conditions and (\ref{constraint2}) are necessary to find the 
points $S_0$ of the stable minimum with
$V(S_0, z_0^a) = 0$, which will be defined by parameters of the 
superpotential $g(S)$ and for good parameter values $<S_R>$ may be 
$<S_R> = 2$, ${\it i.e}$
corresponding to gauge coupling with value 
$\alpha_{GUT} \sim \frac{1}{26}$ 
at the GUT mass scale $M_{GUT} \sim 10^{16}$ Gev. The superpotential 
value in the dilaton direction in this point defines the magnitud of 
the coupling constant $\lambda$ of self-interaction 24 multiplet
\begin{equation}
|g(S_0)| = \lambda.
\label{lam}
\end{equation}
The most direct way to defind $<S_R> = 2$ value through gauge group 
hidden sector in the effective superstring theory is including a  
sector with moduli field direction $T_i$ and the superpotential 
$g_{np}(S,T) = g(S)h(T)$. In this case the K\"ahler potential has the 
form \cite{{ocho},{nueve}}
\begin{equation}
K(S, T, z^a)= -\log(S + \bar S) - 3\log (T + \bar T) + 
\frac{\kappa^2}{2}z^a \bar z^{\bar b}, 
\label{kahler}
\end{equation}
then the constraints (\ref{constraint1},\ref{constraint2}) become 
\begin{equation}
\Pi \equiv G_S G^S + G_T G^T - 1 = 0,\quad |g_{np}(S_0, T_0)|= 
\lambda,
\label{pis}
\end{equation}    
\begin{equation}
\partial_S \Pi = \partial_T \Pi =0, \qquad at \qquad S=S_0, 
\qquad T=T_0,
\label{pos}
\end{equation}
these constraints are imposed only in the hidden sector direction, and
$S_0$, $T_0$ are defined by parameters of the superpotential in the hidden
sector \cite{{ocho},{nueve}}, if we consider the self-dual points of the 
modular space contribution \cite{nueve}, then we have the following 
gravitino mass in the $SU(3)\otimes SU(2)\otimes U(1)$ state
\begin{equation}
m_{3/2} = \left( \frac{5 \pi^{\frac{1}{2}}\lambda}{2^{\frac{3}{2}}} 
\right)^{\sqrt 3} \left(\alpha_{GUT} \right)^{\frac{\sqrt 3}{2}} 
\left(\frac{M_{GUT}}{M_{pl}}\right)^{3\sqrt 3} M_{pl},
\label{massa dos}
\end{equation} 
for $M_{GUT}\sim 10^{16}$ Gev. value and $\alpha_{GUT}\sim \frac{1}{26}$
under $\frac{1}{80}\leq \lambda \leq \frac{1}{20}$ we obtain 
$10^2 \leq m_{3/2} < 10^3$ Gev. Note the circumstance which is due to 
the (\ref{super}); the gravitino mass in the state $SU(4)\otimes U(1)$ is 
related with the gravitino mass in $SU(3)\otimes SU(2)\otimes U(1)$ 
states by relation (for equality values $\alpha_{GUT}$, $\lambda$ 
parameters defined by the hidden sector) 

\begin{equation}
m_{3/2}(SU(4)\otimes U(1))= \left(\frac{10}{27} /5\right)^{\sqrt 3} 
m_{3/2}(SU(3)\otimes SU(2)\otimes U(1)).
\label{massa tres}
\end{equation}  
So, in the case of the hidden sector model with $SU(N)$ gauge group 
the superpotential $g_{np}$ is given by 
$g(S) = -N e^{(-32 \frac{\pi^2}{N} S)}$ and $h(T) = 
(32 \pi e)^{-1}\eta^{-6}(T)$, where $\eta(T)$ is Dedekin eta function 
\cite{{nueve},{diez}}. The coupling constant $\lambda$ in this case is 
defined by 
the magnitud of superpotential $g_{np}(S_0, T_0)$ and can be 
exponentially small, so we cannot excluide from consideration the case 
of arbitrary parameter $\alpha$ values (including $\alpha = 1$ value). 
In that case the constraints (\ref{pis},\ref{pos}) on the hidden sector 
take the forms
\begin{eqnarray}
\Pi(\alpha, S, T) &=& \alpha^2 G_S G^S + \alpha^2 G_T G^T - 3 = 0,
\qquad |g(S_0, T_0)| =\lambda, \\
\partial_S \Pi &=& \partial_T \Pi = 0, \qquad at \qquad\qquad 
S = S_o,\qquad T = T_0. \nonumber 
\label{poso}
\end{eqnarray}  
These constraints give us the stable minimum with vanishing vacuum energy 
(at tree-level) without flat direction and with $<F_S> \ne 0$, 
$<F_T> \ne 0$ we have broken supersymmetry in the moduli direction and 
the broken states SU(5), $<F_{z^a}>=0$. Then, in the state 
with symmetry $SU(3)\otimes SU(2)\otimes U(1)$ of the standard model we 
have the following gravitino mass relation
\begin{equation}
m_{3/2} = \frac{\lambda^{\alpha}} {<T_R>^{\frac{3\alpha}{2}}}
\left(5\pi \alpha_{GUT} \right)^{\frac{\alpha}{2}}
\left(\frac{M_{GUT}}{M_{pl}}\right)^{\alpha} M_{pl},
\label{massa cuatro}
\end{equation}
and the gravitino mass depends on the gauge group of the hidden sector
in the effective supergravity theory. 

\vspace {.5 cm}
{\bf Acknowledgments}: 

We are grateful to I. Bandos, I. Lyanzuridi, L. Marsheva, O. Obreg\'on, 
A. Pashnev and J. Socorro for their interest in this paper. This work was
supported in part by CONACyT grant 3898P-E9608.


\begin{thebibliography}{99} 

\bibitem{uno} S. Ferrara, C. Counnas and F. Zwirner, Nucl. Phys.
{\bf B 429} (1994) 589; 

V.I. Tkach, J. Socorro, J.J. Rosales and J.A. Nieto, submited to 
Class. and Quantum Grav. (1998), hep-th/9807129.     
  
\bibitem{dos} P. Candelas, G. Horowitz, A. Strominger and E. Witten,
Nucl. Phys. {\bf B258} (1985) 46; 

M. Dine, R. Rohm, N. Seiberg and E. Witten, Phys. Lett. {\bf B156} 
(1985) 55; 

J.P. Derendiger, L.E. Iba\~nes and  M.P. Nilles, Phys. Lett. {\bf B155} 
(1985) 65; 

S. Ferrara, C. Kounnas, M. Porrati and F. Zwirner, Nucl. Phys. {\bf B318} 
(1989) 75.

\bibitem{tres} E. Cremmer, B. Julia, J. Scherk, S. Ferrara, L. Girardello
and P. van Nieuwenhuizen, Nucl. Phys. {\bf B147} (1973) 105; 

\bibitem{cuatro} A. M. Chamseddine, R. Arnowitt and P. Nath, Phys. Rev. 
Lett. {\bf 43} (1982) 970.

\bibitem{cinco} V. Kaplunovsky and J. Luis, Phys. Lett. {\bf B306} (1993)
269.

\bibitem{seis} B. de Carlos, J.A. Casas and C. Mu\~nos, Nucl. Phys. 
{\bf B399} (1994) 623; 

C. Kounnas, F. Zwirner and I. Pavel, Phys. Lett. {\bf B335} (1994) 403.

\bibitem{siete} E. Cremmer, S. Ferrara, C. Counnas and D.V. Nanopoulos,
Phys. Lett. {\bf B133} (1983) 61; 

J. Ellis, C. Counnas and D.V. Nonopoulos, Nucl. Phys. {\bf B241} 
(1984) 406; 
Nucl. Phys. {\bf B247} (1984) 373. 

\bibitem{ocho} E. Witten, Phys. Lett. {\bf B155} (1985) 151;
S. Ferrara, C. Counnas, M. Porrati and F. Zwirner, Phys. Lett. {\bf B194}
(1987) 366; 

L. Iba\~nez and D. L\"ust, Nucl. Phys. {\bf B382} (1992) 305;

M. Cveti\v{c}, J. Louis and B. Ovrut, Phys. Lett. {\bf B206} (1988) 227.

\bibitem{nueve} D. L\"ust and T.R. Taylor, Phys. Lett. {\bf B253} (1991)
335; 

A. Font, L.E. Iba\~nez, D. L\"ust and F. Quevedo, Phys. Lett. 
{\bf B245} (1990) 401; 

M. Cveti\v{c}, A. Font, L. Iba\~nez, D. L\"ust and F. Quevedo, Nucl. Phys. 
{\bf B361} (1991) 194.

\bibitem{diez} V. Halyo and E. Halyo, Phys. Lett. {\bf B382} (1996) 89.

\bibitem{once} V.I. Tkach, O. Obreg\'on and J.J. Rosales, Class. and
Quantum Grav. {\bf 13} (1997) 339; 

V.I. Tkach, J.J. Rosales and J. Mart\'{\i}nez, to appear in Class. 
and Quantum Grav. (1998).  

\bibitem{doce} 
V.I. Tkach, J.J. Rosales and J. Socorro, submitted to Mod. Phys. Lett.
{\bf A} (1998), hep-th/9807058; 

V.I. Tkach, J.J. Rosales and J. Socorro, submitted to Phys. Rev. {\bf D} 
(1998), hep-th/980783.

\bibitem{trece} D. Garfinkle, G.T. Horowitz and A. Strominger, Phys. Rev. 
{\bf D43} (1991) 3140; 

A. Mac\'{\i}as and T. Matos, Class. and Quantum Grav. {\bf 13} (1996) 345.

\bibitem{catorce} S. Dimapoulos and M. Georgi, Nucl. Phys. {\bf B193} (1981)
150; 

N. Sakai, Z. Phys. {\bf C11} (1981) 153.

\end{thebibliography}
\end{document}